\documentclass{elsart2}
\journal{Physics Letters {\bf B}}
\usepackage{epsfig}
\usepackage{amssymb}
\usepackage{amstext}
\usepackage{graphicx}

\def\Journal#1#2#3#4{{#1}~{\bf #2} (#4), #3}

\def\NIMA{{\em Nucl.~Instrum.~Methods}~A}
\def\NIMB{{\em Nucl.~Instrum.~Methods}~B}

\def\PLB{{\em Phys.~Lett.}~B}
\def\PRL{\em Phys.~Rev.~Lett.}

\def\EPJ{{\em Eur.~Phys.~J.}~C}

\newcommand{\subrm}[1]{\mbox{\tiny \rm #1}}

\newcommand{\Br}{{\rm Br}}

\newcommand{\reeta}{{\rm{Re(}\eta_{000}\rm{)}}}
\newcommand{\imeta}{{\rm{Im(}\eta_{000}\rm{)}}}

\newcommand{\pid}{\pi^0_{\subrm{Dalitz}}}
\newcommand{\Kbar}{\overline{K^0}}
\newcommand{\KKbar}{K^0 \Kbar}
\newcommand{\kl}{K_{\subrm{L}}}
\newcommand{\ks}{K_{\subrm{S}}}
\newcommand{\kls}{K_{\subrm{L,S}}}

\newcommand{\kpipipi}{K^0 \to 3 \pi^0}
\newcommand{\kspipipi}{\ks \to 3 \pi^0}
\newcommand{\klpipipi}{\kl \to 3 \pi^0}
\newcommand{\klspipipi}{\kls \to 3 \pi^0}
\newcommand{\kpipipid}{K \to \pi^0 \pi^0 \pid}

\newcommand{\klpipipic}{\kl \to \pi^+ \pi^- \pi^0}

\newcommand{\kspipi}{\ks \to \pi^+ \pi^-}

\newcommand{\kspipig}{\ks \to \pi^+ \pi^- \gamma}

\newcommand{\kspizpiz}{\ks \to \pi^0 \pi^0}

\newcommand{\etazzz}{\eta_{000}}

\newcommand{\bdm}{\begin{displaymath}}
\newcommand{\edm}{\end{displaymath}}
\newcommand{\be}{\begin{equation}}
\newcommand{\ee}{\end{equation}}

\begin{document}

\begin{frontmatter}
\title{Search for CP violation in $K^0 \to 3 \pi^0$ decays}
\date{}
%
\author{NA48 collaboration}
\address{\ }
\author{A.~Lai},
\author{D.~Marras}
\address{Dipartimento di Fisica dell'Universit\`a e Sezione dell'INFN di Cagliari, I-09100 Cagliari, Italy} 
\author{J.R.~Batley},
\author{R.S.~Dosanjh\thanksref{threfRanpal}},
\author{T.J.~Gershon\thanksref{threfGershon}},
\author{G.E.~Kalmus},
\author{C.~Lazzeroni},
\author{D.J.~Munday},
\author{E.~Olaiya\thanksref{threfRAL}},
\author{M.A.~Parker},
\author{T.O.~White},
\author{S.A.~Wotton}
\address{Cavendish Laboratory, University of Cambridge, Cambridge, CB3 0HE, U.K.\thanksref{thref3}}
\thanks[threfRanpal]{Present address: Department of Physics, Carleton University, Ottawa, ON, K1S~5B6, Canada}
\thanks[threfGershon]{Present address: High Energy Accelerator Research Organization (KEK), Tsukuba, Japan}
\thanks[threfRAL]{Present address: Rutherford Appleton Laboratory, Chilton, Didcot, Oxon, OX11~0QX, U.K.}
\thanks[thref3]{Funded by the U.K.\ Particle Physics and Astronomy Research Council.}
\author{R.~Arcidiacono\thanksref{threfMIT}},
\author{G.~Barr\thanksref{threfBarr}},
\author{G.~Bocquet},
\author{A.~Ceccucci},
\author{T.~Cuhadar-D\"onszelmann\thanksref{threfTulay}},
\author{D.~Cundy\thanksref{threfZX}},
\author{N.~Doble\thanksref{threfPisa}},
\author{V.~Falaleev},
\author{L.~Gatignon},
\author{A.~Gonidec},
\author{B.~Gorini},
\author{P.~Grafstr\"om},
\author{W.~Kubischta},
\author{A.~Lacourt},
\author{I.~Mikulec\thanksref{threfXY}},
\author{A.~Norton},
\author{B.~Panzer-Steindel},
\author{G.~Tatishvili\thanksref{thref5}\thanksref{threfCMU}},
\author{H.~Wahl\thanksref{threfHW}}
\address{CERN, CH-1211 Gen\`eve 23, Switzerland} 
\thanks[threfMIT]{Present address: Massachusetts Institute of Technology, Cambridge, MA~02139-4307, U.S.A.}
\thanks[threfBarr]{Present address: Department of Physics, University of Oxford, Oxford, OX1~3RH, U.K.}
\thanks[threfTulay]{Present address: University of British Columbia, Vancouver, BC, V6T~1Z1, Canada} 
\thanks[threfZX]{Present address: Istituto di Cosmogeofisica del CNR di Torino, I-10133~Torino, Italy}
\thanks[threfPisa]{Present address: Dipartimento di Fisica, Scuola Normale Superiore e Sezione dell'INFN di Pisa, I-56100~Pisa, Italy}
\thanks[threfXY]{On leave from \"Osterreichische Akademie der Wissenschaften, Institut  f\"ur Hochenergiephysik,  A-1050 Wien, Austria}
\thanks[thref5]{On leave from Joint Institute for Nuclear Research, Dubna, 141980, Russian Federation}
\thanks[threfCMU]{Present address: Carnegie Mellon University, Pittsburgh, PE~15213, U.S.A.}
\thanks[threfHW]{Present address: Dipartimento di Fisica dell'Universit\`a e Sezione dell'INFN di Ferrara, I-44100~Ferrara, Italy}
\author{C.~Cheshkov\thanksref{threfCERN}},
\author{P.~Hristov\thanksref{threfCERN}},
\author{V.~Kekelidze},
\author{D.~Madigojine},
\author{N.~Molokanova},
\author{Yu.~Potrebenikov},
\author{A.~Zinchenko}
\address{Joint Institute for Nuclear Research, Dubna, Russian    Federation}  
\thanks[threfCERN]{Present address: CERN, CH-1211 Gen\`eve~23, Switzerland}
%
%
\author{V.~Martin\thanksref{threfNW}},
\author{P.~Rubin\thanksref{threfPhil}},
\author{R.~Sacco\thanksref{threfSacco}},
\author{A.~Walker}
\address{Department of Physics and Astronomy, University of Edinburgh, JCMB King's Buildings, Mayfield Road, Edinburgh, EH9~3JZ, U.K.}
\thanks[threfNW]{Present address: Northwestern University, Department of Physics and Astronomy, Evanston, IL~60208, U.S.A.}
\thanks[threfPhil]{Present address: Department of Physics and Astronomy, George Mason University, Fairfax, VA~22030, U.S.A.;
funded by US NSF under grants 9971970 and 0140230.} 
\thanks[threfSacco]{Present address: Department of Physics, Queen Mary University, London, E1~4NS, U.K.}
%
%
\author{M.~Contalbrigo},
\author{P.~Dalpiaz},
\author{J.~Duclos},
\author{M.~Fiorini},
\author{P.L.~Frabetti\thanksref{threfFrabetti}},
\author{A.~Gianoli},
\author{M.~Martini},
\author{F.~Petrucci},
\author{M.~Savri\'e}
\address{Dipartimento di Fisica dell'Universit\`a e Sezione    dell'INFN di Ferrara, I-44100 Ferrara, Italy}
\thanks[threfFrabetti]{Present address: Joint Institute for Nuclear Research, Dubna, 141980, Russian Federation}
%
%
\author{A.~Bizzeti\thanksref{threfXX}},
\author{M.~Calvetti},
\author{G.~Collazuol\thanksref{threfPisa}},
\author{G.~Graziani\thanksref{threfGG}},
\author{E.~Iacopini},
\author{M.~Lenti},
\author{F.~Martelli\thanksref{thref7}},
\author{M.~Veltri\thanksref{thref7}}
\address{Dipartimento di Fisica dell'Universit\`a e Sezione dell'INFN di Firenze, I-50125~Firenze, Italy}
\thanks[threfXX]{Dipartimento di Fisica dell'Universit\`a di Modena e Reggio Emilia, I-41100~Modena, Italy}
\thanks[threfGG]{Present address: DSM/DAPNIA - CEA Saclay, F-91191 Gif-sur-Yvette, France}
\thanks[thref7]{Istituto di Fisica dell'Universit\`a di Urbino, I-61029~Urbino, Italy}
%
%
\author{K.~Eppard},
\author{M.~Eppard\thanksref{threfCERN}},
\author{A.~Hirstius\thanksref{threfCERN}},
\author{K.~Kleinknecht},
\author{U.~Koch},
\author{L.~K\"opke},
\author{P.~Lopes da Silva}, 
\author{P.~Marouelli},
\author{I.~Mestvirishvili},
\author{I.~Pellmann\thanksref{threfDESY}},
\author{A.~Peters\thanksref{threfCERN}},
\author{S.A.~Schmidt},
\author{V.~Sch\"onharting},
\author{Y.~Schu\'e},
\author{R.~Wanke\corauthref{cor}},
\author{A.~Winhart},
\author{M.~Wittgen\thanksref{threfSLAC}}
\address{Institut f\"ur Physik, Universit\"at Mainz, D-55099~Mainz, Germany\thanksref{thref6}}
\thanks[threfDESY]{Present address: DESY Hamburg, D-22607~Hamburg, Germany}
\corauth[cor]{Corresponding author.{\em Email address:} Rainer.Wanke@uni-mainz.de}
\thanks[threfSLAC]{Present address: SLAC, Stanford, CA~94025, U.S.A.}
\thanks[thref6]{Funded by the German Federal Minister for Research and Technology (BMBF) under contract 7MZ18P(4)-TP2}
\author{J.C.~Chollet},
\author{L.~Fayard},
\author{L.~Iconomidou-Fayard},
\author{G.~Unal},
\author{I.~Wingerter-Seez}
\address{Laboratoire de l'Acc\'el\'erateur Lin\'eaire, IN2P3-CNRS, Universit\'e de Paris-Sud, 91898~Orsay, France\thanksref{threfOrsay}}
\thanks[threfOrsay]{Funded by Institut National de Physique des Particules et de Physique Nucl\'eaire (IN2P3), France}
\author{G.~Anzivino},
\author{P.~Cenci},
\author{E.~Imbergamo},
\author{P.~Lubrano},
\author{A.~Mestvirishvili\thanksref{threfAlexi}},
\author{A.~Nappi},
\author{M.~Pepe},
\author{M.~Piccini}
\address{Dipartimento di Fisica dell'Universit\`a e Sezione dell'INFN di Perugia, I-06100~Perugia, Italy}
\thanks[threfAlexi]{Present address: Department of Physics and Astronomy, University of Iowa, Iowa City, IA~52242-1479, U.S.A.}
%
%
\author{R.~Casali},
\author{C.~Cerri},
\author{M.~Cirilli\thanksref{threfCERN}},
\author{F.~Costantini},
\author{R.~Fantechi},
\author{L.~Fiorini},
\author{S.~Giudici},
\author{I.~Mannelli},
\author{G.~Pierazzini},
\author{M.~Sozzi}
\address{Dipartimento di Fisica, Scuola Normale Superiore e Sezione dell'INFN di Pisa, I-56100 Pisa, Italy} 
%
%
\author{J.B.~Cheze},
\author{M.~De Beer},
\author{P.~Debu},
\author{F.~Derue\thanksref{threfDerue}},
\author{A.~Formica},
\author{R.~Granier de Cassagnac\thanksref{threfGranier}},
\author{G.~Gouge},
\author{G.~Marel},
\author{E.~Mazzucato},
\author{B.~Peyaud},
\author{R.~Turlay\thanksref{Deceased}},
\author{B.~Vallage}
\address{DSM/DAPNIA - CEA Saclay, F-91191~Gif-sur-Yvette, France} 
\thanks[threfDerue]{Present address: Laboratoire de l'Acc\'el\'erateur Lin\'eaire, IN2P3-CNRS, Universit\'e de Paris-Sud, 91898~Orsay, France}
\thanks[threfGranier]{Present address: Laboratoire Leprince-Ringuet, Ecole polytechnique/IN2P3, 91128~Palaiseau, France}
\thanks[Deceased]{Deceased.}
\author{M.~Holder},
\author{A.~Maier\thanksref{threfCERN}},
\author{M.~Ziolkowski }
\address{Fachbereich Physik, Universit\"at Siegen, D-57068 Siegen, Germany\thanksref{thref8}}
\thanks[thref8]{Funded by the German Federal Minister for Research and Technology (BMBF) under contract 056SI74}
\newpage
\author{C.~Biino},
\author{N.~Cartiglia},
\author{F.~Marchetto}, 
\author{E.~Menichetti},
\author{N.~Pastrone}
\address{Dipartimento di Fisica Sperimentale dell'Universit\`a e Sezione dell'INFN di Torino, I-10125~Torino, Italy} 
\author{J.~Nassalski},
\author{E.~Rondio},
\author{M.~Szleper\thanksref{threfNW}},
\author{W.~Wislicki},
\author{S.~Wronka}
\address{Soltan Institute for Nuclear Studies, Laboratory for High Energy Physics, PL-00-681~Warsaw, Poland\thanksref{thref9}}
\thanks[thref9]{Supported by the Committee for Scientific Research grants 5P03B10120, SPUB-M/CERN/P03/DZ210/2000 and SPB/CERN/P03/DZ146/2002}
\author{H.~Dibon},
\author{M.~Jeitler},
\author{M.~Markytan},
\author{G.~Neuhofer},
\author{M.~Pernicka},
\author{A.~Taurok},
\author{L.~Widhalm}
\address{\"Osterreichische Akademie der Wissenschaften, Institut  f\"ur Hochenergiephysik, A-1050~Wien, Austria\thanksref{thref10}}
\thanks[thref10]{Funded by the Austrian Ministry for Traffic and Research under the contract GZ 616.360/2-IV GZ 616.363/2-VIII, and by the Fonds f\"ur Wissenschaft und Forschung FWF Nr.~P08929-PHY}

\vspace*{\fill}
\begin{abstract}
Using data taken during the year 2000 with the NA48 detector at the CERN SPS, a search for the
CP violating decay $\kspipipi$ has been performed. 
From a fit to the lifetime distribution of about $4.9$~million
reconstructed $K^0/\overline{K^0} \to 3 \pi^0$ decays,
the CP violating amplitude $\etazzz = A(\kspipipi)/A(\klpipipi)$
has been found to be
$\reeta = -0.002 \pm 0.011 \pm 0.015$ and $\imeta = -0.003 \pm 0.013 \pm 0.017$.
This corresponds to an upper limit on the branching fraction
of $\Br(\kspipipi) < 7.4 \times 10^{-7}$ at 90\% confidence level.
The result is used to improve knowledge of ${\rm Re}(\epsilon)$
and the CPT violating quantity ${\rm Im}(\delta)$
via the Bell-Steinberger relation.
\end{abstract}

\end{frontmatter}

\setcounter{footnote}{0}

\newcommand{\spaceafterfloat}{\vspace*{3mm}} 

%
%

\section{Introduction}

\vspace*{-3mm}

The violation of CP symmetry 
was discovered in 1964 in the decay of
the long-lived $\kl$ meson to two charged pions~\cite{bib:ccft}.
Since then, other CP violating $\kl$ decay modes --- in particular direct CP violation --- and CP violation in $B^0$ decays
have been observed.
The short-lived neutral kaon $\ks$, too, should manifest
CP violating decay amplitudes.
However, due to the large width of the $\ks$ meson, the branching ratios of
CP violating $\ks$ decays are 6 orders of magnitude smaller than
the branching ratios of the main CP violating $\kl$ decays.
One unambiguous signature of CP violation  in $\ks$ decays 
would be the observation of the decay $\kspipipi$.
Within the Standard Model its branching ratio is predicted to be $1.9 \times 10^{-9}$.

Since the kaon is spinless, the $3 \pi^0$ final state has
a well-defined CP eigenvalue of $-1$, when neglecting direct CP violation,
and the decay $\kspipipi$ is CP forbidden.
The CP violating parameter $\etazzz$ is defined as
the amplitude ratio
\begin{equation}
\label{eqn:eta000}
\etazzz \equiv \frac{A(\kspipipi)}{A(\klpipipi)}.
\end{equation}
When assuming CPT conservation and neglecting
isospin $I=3$ and non-symmetric $I=1$ final states,
$\etazzz = \epsilon + i \: {\rm Im}(A_1) / {\rm Re}(A_1)$
with the parameter $\epsilon$ of indirect CP violation and the $I=1$ amplitude $A_1$. 
The imaginary part of $\etazzz$ is in principle sensitive to direct CP violation~\cite{bib:pdg}.

Additional interest in $\kspipipi$ decays
arises from the search for CPT violation.
The Bell-Steinberger relation~\cite{bib:bellsteinberger}
links possible CPT violation in the $\KKbar$ mixing matrix with CP violating
amplitudes in $\kl$ and $\ks$ decays via the conservation of probability.
At present, the limit on CPT violation is limited by poor knowledge of $\etazzz$.

In a recent investigation,
the CPLEAR experiment found
$\reeta = 0.18 \pm 0.15$ and $\imeta = 0.15 \pm 0.20$
in the decay of flavour-tagged $K^0$ and $\Kbar$ mesons~\cite{bib:ks3pi0_cplear},
which corresponds to
an upper bound of Br($\kspipipi) < 1.9 \times 10^{-5}$ at $90\%$
confidence level. In addition, the SND collaboration has 
set a limit of Br($\kspipipi) < 1.4 \times 10^{-5}$~\cite{bib:ks3pi0_snd}.

In this Letter we report an improved measurement of the $\etazzz$ parameter
using data collected from a short neutral beam with the NA48 detector at CERN.
Sensitivity to $\etazzz$ comes from $\ks/\kl \to 3 \pi^0$ interference at small decay times near
the target. A beam of pure $\kl$ decays, taken under the same experimental conditions, was used
for normalization.
The experimental set-up is described in the next section,
while section \ref{sec:selection} deals with event selection and reconstruction.
The fit to extract the parameter $\etazzz$ is described in Section
\ref{sec:dataanalysis}. 
The last section discusses the implications of the result with respect to the $\kspipipi$ branching fraction
and the limit on CPT violation.

\section{Experimental Setup}
\label{sec:setup}

\vspace*{-3mm}

\begin{figure}
\begin{center}
\epsfig{file=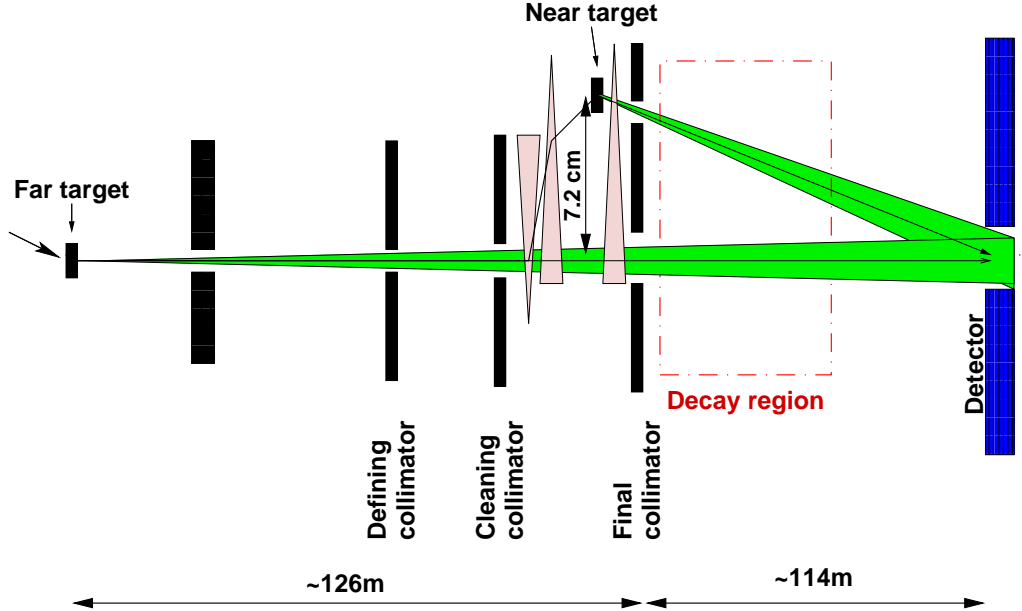,width=0.9\linewidth}
\caption{Beam-line of the NA48 experiment during the 2000 data taking.}
\label{fig:beamline}
\end{center}
\spaceafterfloat
\end{figure}

The NA48 experiment was designed for the measurement of direct
CP violation in neutral kaon decays.
Its main feature is two simultaneous, almost collinear 
beams of neutral kaons derived from proton beams from the CERN SPS delivered to
two fixed targets~\cite{bib:na48_beams}. 
The kaon beams have a common decay region and decays from both beams are recorded with the same detector (see Fig.~\ref{fig:beamline}).
Both targets are made from beryllium and have a length of $400$~mm and a diameter of $2$~mm.
The far target is located $126$~m before the beginning of the decay region
while the near target is only $6$~m up-stream of the fiducial region
and displaced by $7.2$~cm in the vertical direction from the axis of the far-target beam.   
The two beam axes have an angle of $0.6$~mrad with respect to each other 
and cross at the longitudinal position
of the electromagnetic calorimeter, $120$~m down-stream of the near target.
In both beams, charged particles are deflected by sweeping magnets.
Particle decays from the far target are almost exclusively $\kl$ decays,
while decays originating from the near target are mainly $\ks$ (and neutral hyperon) decays
with, however, a small component of $\kl$ decays. 

The analysis reported here is based on
data from a 40-day run period in 2000, with only the near target in operation.
The neutral beam was produced by $400$~GeV/$c$ protons at
a production angle of $3.0$~mrad, with a beam intensity of about $10^{10}$ 
protons during a $3.2$~s long SPS spill. This was about a factor of 300 higher
than the typical intensity of the near-target beam during the direct CP violation measurement.
This high-intensity run period was preceded by a 30-day run with only a far-target 
beam under the same beam conditions, except for the 
proton beam energy being $450$~GeV and the production angle being $2.4$~mrad.
The data from this first run period were used for normalization purposes.

In the usual configuration, 
the main NA48 detector elements are a magnetic spectrometer for reconstruction of charged
particles, followed by a hodoscope for charged particles and a 
liquid krypton electromagnetic calorimeter (LKr).
During both run periods in 2000, the spectrometer drift chambers were absent and 
its vessel evacuated, leaving no material
between the final collimator and the hodoscope for charged particles
directly in front of the calorimeter.
In the far-target run period, the spectrometer magnet was powered with an integrated field of 0.88~Tm in order to
have the same running conditions for systematic studies of the direct CP violation measurement.
In the subsequent near-target run period the magnet was off.

The liquid krypton calorimeter measures the 
energies, positions, and times of electromagnetic showers initiated by photons and electrons~\cite{bib:lkr}.
It has a length of 127~cm, corresponding to 27 radiation lengths,
and consists of 13212 cells in a projective tower geometry
which points to the middle of the decay volume.
The active volume has an octagonal shaped cross section of about 5.5~m$^2$.
Each cell has a $2 \times 2$~cm$^2$ cross section and is formed by copper-beryllium
ribbons which are extended longitudinally in a $\pm 48$~mrad accordion structure.
The cells are contained in a cryostat
filled with about 10~m$^3$ of liquid krypton at a temperature of 120~K.
The initial ionization signal induced on the electrodes
is amplified, shaped, and digitized by 40~MHz FADCs.
The energy resolution of the calorimeter was
\begin{equation}
\frac{\sigma(E)}{E} \; \simeq \; \frac{0.090}{E} \; \oplus \; \frac{0.032}{\sqrt{E}} \; \oplus \; 0.0042
\end{equation}
with $E$ in GeV. The spatial and time resolutions were better than 1.3~mm and 300~ps, respectively, for a photon
with energy above 20~GeV.
The read-out system was calibrated by a charge pulse every burst during data taking.
The relative calibration of the individual cells was determined
by using $K_{e3}$ decays during the 1998 run period and checked to be similar
in 2000 by $\pi^0/\eta \to \gamma \gamma$ decays produced in thin plastic targets
in a special run with a $\pi^-$ beam.  
The final calibration of the overall energy scale was performed 
by fitting the effective edge of the collimator as reconstructed in the data to the 
one in the Monte Carlo simulation.

The hadron calorimeter follows the LKr calorimeter and was used as a veto-counter.
It is composed of forty-eight 24~mm thick steel plates interleaved with
scintillator plates and measures energies and horizontal and vertical positions of hadronic showers.
A more complete description of the NA48 detector can be found in Ref.~\cite{bib:epsprime01}.

The trigger decision for $3\pi^0$ events was based on projections of the deposited energy 
in the liquid krypton calorimeter~\cite{bib:lkrtrigger}.
In both run periods the trigger required a total deposited energy of at least 50~GeV.
In addition, the radius of the energy centre-of-gravity had to be less than 15~cm
from the detector axis and the proper kaon lifetime, measured from the final collimator, had to be less than 9~$\ks$ lifetimes
with both radius and proper lifetime computed online from the moments of the energy depositions in the calorimeter.
The trigger efficiency was determined using a $3\pi^0$ data sample triggered
by a scintillating fiber hodoscope located inside the calorimeter.
It was measured to be $99.8\%$
in both the near- and the far-target run periods, 
and showed no dependence on energy or decay vertex position within the decay volume used in the analysis.

\section{Event Selection and Reconstruction}
\label{sec:selection}

\vspace*{-3mm}

To identify $K^0 \to 3 \pi^0 \to 6 \gamma$ events and to determine their kinematics,
the measured energies $E_i$, positions $x_i$ and $y_i$, and times of the photon showers in the liquid krypton 
electromagnetic calorimeter are used.
From the energies deposited in the LKr cells, clusters were formed, which had to fulfill the 
following selection criteria.
The cluster energies were required to be above 3~GeV and below 100~GeV, within
the range of the linear energy response of the calorimeter.
To avoid energy losses, each cluster had to be more than 5~cm
from the edge of the beam pipe and from the outer edge of the sensitive area.
In addition, the distance to the closest dead cell of the calorimeter
was required to be larger than 2~cm.

On all combinations of 6 clusters which passed these requirements, the following
further selection criteria were applied.
To avoid difficulties with the sharing of energy among close clusters,
the minimum distance between two clusters had to be at least 10~cm.
All six clusters were required to lie within 2~ns of the average cluster time.
The sum of cluster energies had to lie above 60~GeV and below 185~GeV.
The radial position $r_{\subrm{cog}}$ of the energy 
centre-of-gravity at the calorimeter
had to be less than 7~cm in the near-target and less than 4~cm in the far-target run, which had stronger beam collimation.

To avoid background from accidental pile-up,
events with more than one such combination of clusters and 
those with additional clusters with an energy above 1.5~GeV within 3~ns of the
event time were rejected.

From the surviving candidates, the longitudinal vertex along the
$z$ direction was reconstructed from all pairings by assuming 
the decay of a $K^0$ with the nominal mass $m_K$:
\begin{equation}
\label{eqn:zvertex}
z_{\subrm{vertex}} \; = \; z_{\subrm{LKr}}
                        - \frac{1}{m_K} \sqrt{\sum_{i=1}^6 \sum_{j>i}^6 E_i E_j \left[ (x_i - x_j )^2 + (y_i - y_j )^2 \right] }
\end{equation}
with the distance $z_{\subrm{LKr}}$ between the near target and the calorimeter.
The resolution of the reconstructed longitudinal vertex position is about 60~cm, corresponding
to $\sim 0.1$~$\ks$~lifetimes for typical kaon energies. 

Using the longitudinal vertex position $z_{\subrm{vertex}}$, 
the invariant two-photon masses $m_1$, $m_2$, and $m_3$ were computed for 
all 15 possible photon pairing combinations, and a $\chi^2$-like variable
was constructed as
\begin{equation}
\! \! \! \chi^2_{3\pi^0}  =    \left( \frac{\frac{1}{3}(m_1 + m_2 + m_3) - m_{\pi^0}}{\sigma_1} \right)^2
           + \left( \frac{\frac{1}{2}(m_1 - \frac{m_2 + m_3}{2})}{\sigma_2} \right)^2
           + \left( \frac{\frac{1}{2}(m_2 - m_3)}{\sigma_3} \right)^2
\end{equation}
where the resolutions $\sigma_i$,
parameterized as a function of the smallest photon energy,
were determined from data.
The combination with the lowest value of $\chi^2_{3\pi^0}$ was chosen. 
In addition, a value of $\chi^2_{3\pi^0}$ less than $90$ was required for this combination.

\begin{figure}
\begin{center}
\epsfig{file=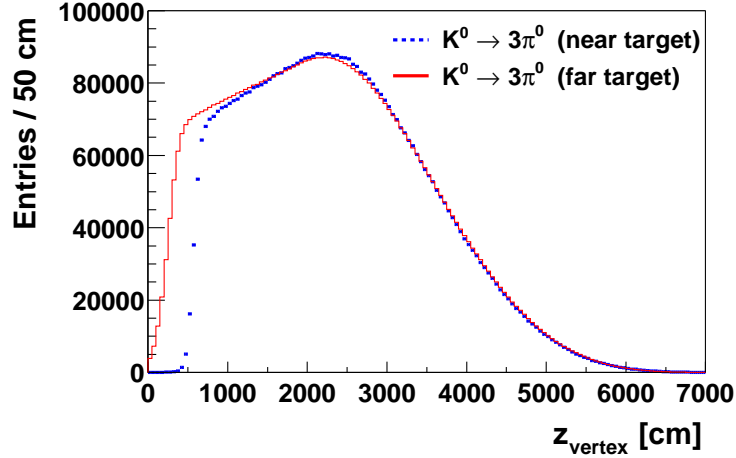,width=0.65\linewidth}
\caption{Longitudinal vertex position of reconstructed
$\kpipipi$ events. In the figure, the far-target data have 
been normalized to the near-target data in the fiducial decay region.}
\label{fig:zdist}
\end{center}
\spaceafterfloat
\end{figure}

To reject possible background from hadrons, we required
the total energy deposited in the hadron calorimeter within 15~ns of the event time 
to be less than 3~GeV, which removed $0.15\%$ of the signal events.

\begin{figure}
\begin{center}
\epsfig{file=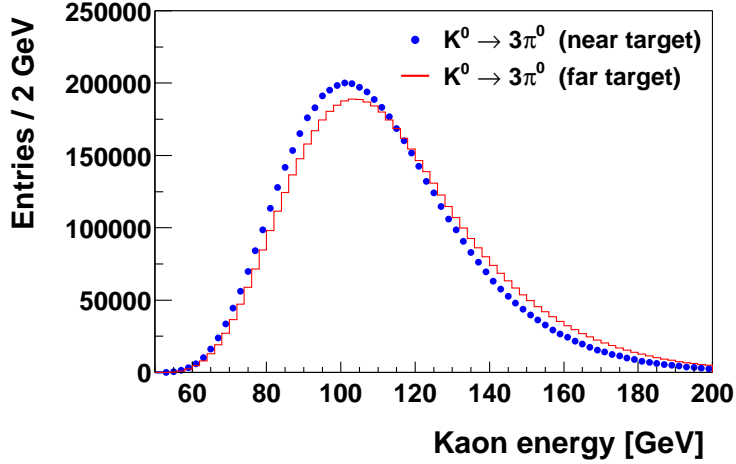,width=0.65\linewidth}
\caption{Energy spectra of reconstructed
$\kpipipi$ events. The far-target data have been normalized to the near-target data.}
\label{fig:energy}
\end{center}
\spaceafterfloat
\end{figure}
\begin{figure}[t]
\begin{center}
\epsfig{file=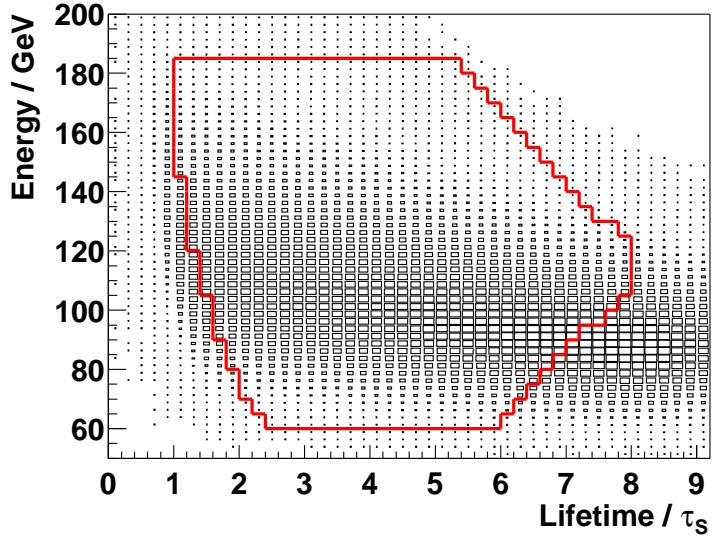,width=0.65\linewidth}
\caption{Energy versus lifetime, measured from the target, of accepted $K\to 3 \pi^0$ events from
the near-target run. The contour encloses the accepted region for the $\etazzz$ fit.}
\label{fig:EKtau}
\end{center}
\spaceafterfloat
\end{figure}

The $z_{\subrm{vertex}}$ and energy distributions of the
selected events are shown in Figs.~\ref{fig:zdist} and \ref{fig:energy}.
For the fit of the parameter $\etazzz$, only events with a 
longitudinal vertex position $z_{\subrm{vertex}} > 8$~m were considered
to avoid detector resolution effects for vertex positions near the final collimator at $z = 6$~m.
The down-stream vertex region was limited by $z_{\subrm{vertex}} < 55$~m 
and a maximum lifetime of $8 \, \tau_{K_S}$.
In addition, events with lifetimes $t_{\subrm{coll}}/\tau_{K_S} >  0.8 + 0.06 \times E_K / {\rm GeV} $,
measured from the final collimator, were rejected to avoid a region at low energies and high lifetimes
where the trigger was partly inefficient.
The accepted region in proper time and energy 
is shown in Fig.~\ref{fig:EKtau}.

In total, about $4.9 \times 10^6$ $\klspipipi$ events were reconstructed
from the data of the near-target run and about $109 \times 10^6$ $\klpipipi$ events
from the far-target run data. 


\section{Data Analysis}
\label{sec:dataanalysis}. 

\vspace*{-18mm}

\subsection{Method of the Measurement}
\label{sec:method}

\vspace*{-3mm}

\begin{figure}
\begin{center}
\epsfig{file=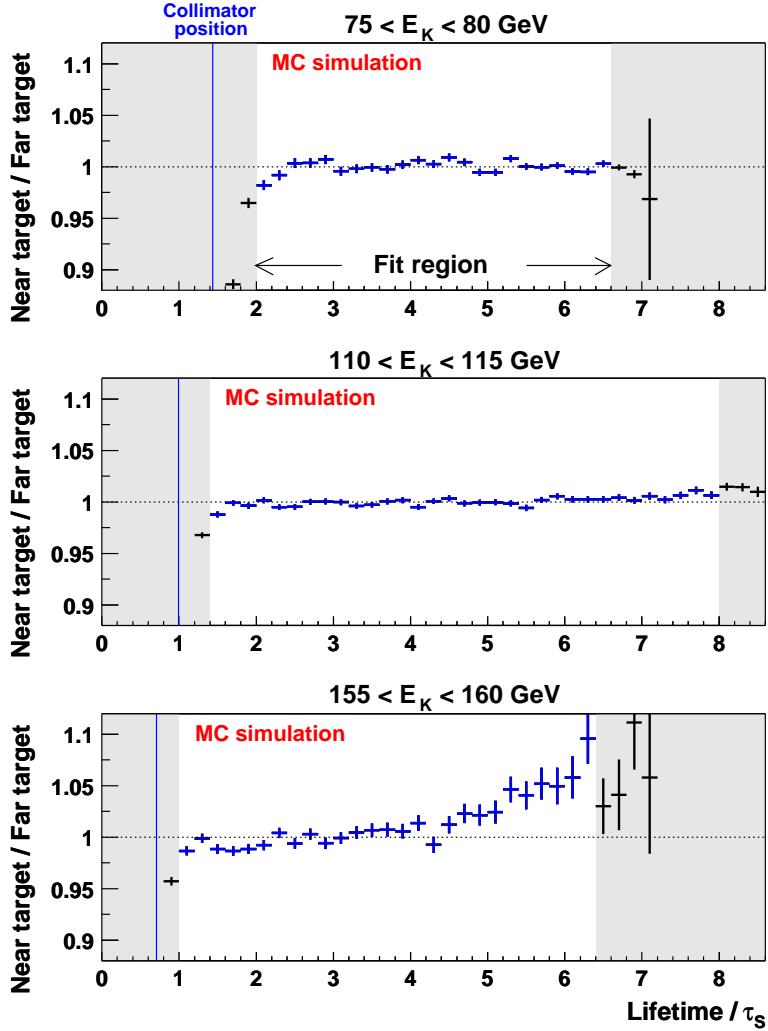,width=0.7\linewidth}
\caption{Ratio $\epsilon^{\subrm{near}} / \epsilon^{\subrm{far}}$ of the acceptances 
for $3 \pi^0$ events from the two kaon beams in three kaon energy ranges, as determined by Monte Carlo simulation.}
\label{fig:geocorrection}
\end{center}
\spaceafterfloat
\end{figure}

At the targets, $\kl$ and $\ks$ mesons are produced by strong interactions in equal amounts. 
The $K \to 3 \pi^0$ intensity as a function of proper time $t$, measured from the target, is then given by
\begin{equation}
\begin{array}{rcl}
I_{3\pi^0}(t) & \; \propto \; &   e^{- \Gamma_L \, t}
                          \; + \; |\etazzz|^2 \, e^{- \Gamma_S \, t} \\
              &         & + \; 2 \: D(p) \, \left( \reeta \cos (\Delta m \, t)  - \imeta \sin (\Delta m \, t) \right) \\
              &         &  \multicolumn{1}{r}{\times \, e^{ - \frac{1}{2}(\Gamma_S + \Gamma_L) \, t}}
\end{array}
\label{eqn:intensity}
\end{equation}
with the total $\kl$ and $\ks$ widths $\Gamma_L$ and $\Gamma_S$ and the $\kl \ks$ mass difference $\Delta m$.
The {\em dilution} $D(p) = (N_{K^0} - N_{\Kbar})/(N_{K^0} + N_{\Kbar})$ describes
the momentum dependent production asymmetry between $K^0$ and $\Kbar$ at the target.
For the $\etazzz$ measurement we analyzed data from the near-target run period,
using the pure $\klpipipi$ from the immediately
preceding far-target run period to correct for trigger, acceptance, and reconstruction efficiencies.
The difference of the kaon beam momentum spectra between the two periods
was taken into account by performing the analysis in 5~GeV wide bins of energy 
covering a range from 60 to 185~GeV.
The two set-ups had small differences in geometry, incident kaon beam angles, and collimation.
Based on Monte Carlo studies with samples of about 90~million reconstructed events for each beam,
the ratio
\begin{equation}
\label{eqn:fit_func}
f_{3\pi^0}(t) = \frac{N^{\subrm{near}}(t)}{N^{\subrm{far}}(t)} \; \Bigg{/} \;
                \frac{\epsilon^{\subrm{near}}(t)}{\epsilon^{\subrm{far}}(t)}
\end{equation}
with the numbers $N^{\subrm{near}}$ and $N^{\subrm{far}}$ of reconstructed events and 
the acceptances $\epsilon^{\subrm{near}}$ and $\epsilon^{\subrm{far}}$ for near and far targets was determined.
The geometrical correction $\epsilon^{\subrm{near}}(t)/\epsilon^{\subrm{far}}(t)$
between the two beams nearly equals to 1 and is shown in Fig.~\ref{fig:geocorrection} for three different
energy intervals.
Neglecting this correction would shift the $\etazzz$ result
by $\Delta \reeta = -0.03$ and  $\Delta \imeta = 0.03$.

The intensity $I_{3\pi^0}(t)$ as function of proper time might be altered if
there is significant $\ks$ regeneration by $\kl$ mesons hitting the final collimator.
This possibility has been considered and found to be negligible for the measurement presented here.

In the near-target data decays
$K \to \pi^0 \pi^0 \pi^0_D$ with one pion undergoing a Dalitz decay
$\pi^0_D \to e^+ e^- \gamma$ had to be taken into account.
Since one shower could have missed the detector or could have had an energy
below the detection limit of 1.5~GeV, these decays could be mis-identified
as $K \to 3\pi^0 \to 6 \gamma$ events.
In many such cases, due to the energy loss, the decay vertex position $z_{\subrm{vertex}}$
and the kaon lifetime were reconstructed further down-stream (see Eq.~\ref{eqn:zvertex}).
For the near-target run the
mis-identification rate with respect to the acceptance of good
$3 \pi^0$ events was $\epsilon(2 \pi^0 \pi^0_D)/\epsilon(3 \pi^0) \approx 40\%$,
while it was negligible in the far-target run due 
to the presence of the magnetic field.
Dalitz decays were taken into account in the
Monte Carlo generation of the near-target $\klpipipi$ events.

\subsection{Fit to the data}
\label{sec:fit}

\vspace*{-3mm}

For all $\kpipipi$ candidates, the proper lifetime of the kaon,
measured from the position of the near target at $z=0$, 
is computed as $t = z_{\subrm{vertex}} / (\gamma \beta c ) \approx z_{\subrm{vertex}} \, m_K / (E_K c)$,
where $m_K$ is the kaon mass taken from the PDG and $E_K$ is the sum of the cluster energies.



A global least-squares fit to the time evolution $I_{3\pi^0}(t)$ (Eq.~\ref{eqn:intensity})
is performed
on the corrected time distributions $f_{3\pi^0}(t)$ (Eq.~\ref{eqn:fit_func}) for each kaon energy interval.
Free parameters in the fit are $\reeta$, $\imeta$,
and the normalization constants of each energy interval.
The $\KKbar$ dilution is assumed to be linearly dependent on energy, 
as expected from the parton model.
It is obtained using data from the NA31 experiment~\cite{bib:na31_dilution},
corrected for the different momentum spectrum and production angle~\cite{bib:atherton}
and parameterized as a linear function of the energy.

\begin{figure}[t]
\begin{center}
\epsfig{file=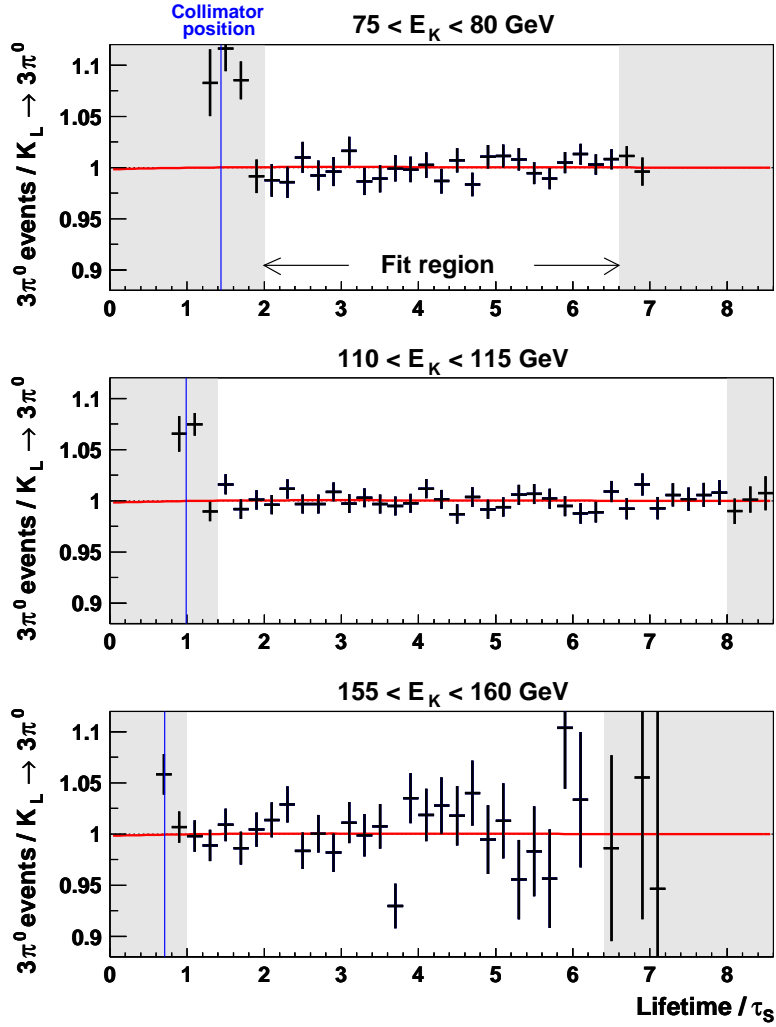,width=0.7\linewidth}
\caption{The ratio of near-target over far-target $3\pi^0$ data, corrected for
acceptance differences, for three different energy intervals.
The points with error bars are the data.
The continuous curve, which practically coincides with 1, shows the result of the fit for $\etazzz$.}
\label{fig:fitresult}
\end{center}
\spaceafterfloat
\end{figure}

The result of the fit gives $\reeta = -0.002 \pm 0.011$
and $\imeta = -0.003 \pm 0.013$ with a correlation coefficient of $0.77$ 
between real and imaginary parts (Fig.~\ref{fig:fitresult}). The $\chi^2$ of the fit is $689$ with $660$ degrees of freedom.

\subsection{Systematic uncertainties}
\label{sec:systematics}

\vspace*{-3mm}

We have investigated contributions to the systematic uncertainties
from detector acceptance, from calorimeter energy scale and non-linearities,
from possible backgrounds and accidental activity, and from $\KKbar$ dilution.

As described above, the acceptance determination relied on data from the far-target run,
with the Monte Carlo simulation used to calculate residual geometrical differences.
Many checks of the acceptance calculation were performed.
Differences in the reconstruction efficiency between the two runs were investigated
by varying the $\chi^2_{3\pi^0}$ cut, which led to
uncertainties of $\Delta \reeta = \pm 0.009$ and $\Delta \imeta = \pm 0.013$.
Remaining resolution effects near the collimator were studied by varying the up-stream vertex cut and gave an uncertainty
of $\Delta \reeta = \pm 0.010$, while the imaginary part --- playing a role only at larger lifetimes --- was unaffected.
By comparing two Monte Carlo samples with different target-collimator geometries, 
we estimated $\Delta \reeta = \pm 0.005$ and $\Delta \imeta = \pm 0.007$
as uncertainties related to the $z_{\subrm{vertex}}$ dependence of the acceptance.

While the run-to-run variation of the calorimeter energy scale 
was smaller than $10^{-4}$, the overall energy scale is only known to
$10^{-3}$ in the 2000 run period\footnote{In the usual configuration for the direct CP violation measurement,
the beginning of the decay volume was defined by a scintillating anti-counter in the near-target beam.
In 2000, this anti-counter was removed to be able to take a 300 times more intense neutral beam, and the reconstructed
collimator edge was used as reference position.}.
By varying the energy scale within $\pm 10^{-3}$ we estimated systematic uncertainties of
$\Delta \reeta \simeq \Delta \imeta \simeq \pm 0.001$.
Furthermore, we investigated the effect of possible non-linearities in the shower energy reconstruction
by modifying the reconstructed energy by an amount
$\Delta E/E = \alpha/E +  \beta E + \gamma r$,
with $r$ being the radial distance of the shower from the central detector axis.
The allowed ranges of the parameters $\alpha$, $\beta$, and $\gamma$ 
were determined from studies of $K_{e3}$, $K \to \pi^0 \pi^0$, $\kpipipi$, and
$\pi^0 / \eta \to \gamma \gamma$ decays from older data collected in 1999, and found to be
$\alpha = \pm 10$~MeV, $\beta = \pm 2 \times 10^{-5}$~GeV$^{-1}$, and  $\gamma = \pm 10^{-5}$~cm$^{-1}$.
When varying the constants within these ranges,
the fitted value of $\etazzz$ varied by $\Delta \reeta = \pm 0.001$ and $\Delta \imeta = \pm 0.002$.

Uncertainties in the acceptance of $\kpipipid$ decays in the near-target run result in an uncertainty of $\Delta \imeta = \pm 0.001$,
while the uncertainty on $\reeta$ is less than 0.001.

Background processes which could fake a $\kpipipi$ event play a role
only in the near-target run, where the majority of decays comes from $\ks$ mesons and
$\Lambda$ and $\Xi^0$ hyperons.
To estimate the possible background in the $\kpipipi$ near-target data sample,
we compared the tails of the $r_{\subrm{cog}}$ distributions of $\kpipipi$ events
and $\kspizpiz$ events, which are practically background-free, and found no significant discrepancy.
Pile-up events could either fake $\kpipipi$ events or lead to accidental losses due 
to the cut on the number of clusters in the calorimeter.
From studying time side-bands and cluster widths, we found no indication for pile-up events.
A conservative upper limit on the effect of possible background and accidental losses in the near-target beam was estimated
by loosening the cuts on the energy in the hadron calorimeter and the number of clusters
in the LKr calorimeter, from which we found $\Delta \reeta = \pm 0.002$ and  $\Delta \imeta = \pm 0.009$.

As described in the previous section, the $\KKbar$ dilution was taken from measured data of the NA31 experiment.
By varying the dilution within the measurement errors, we obtained 
an uncertainty on our fit result of $|\Delta \reeta| = |\Delta \imeta| \le 0.001$.


The systematic uncertainties for $\etazzz$ are summarized
in Table~\ref{tab:systematics}. The individual contributions were added in quadrature, taking
correlations between $\reeta$ and $\imeta$ into account.

\begin{table}[t]
\begin{center}
\begin{tabular}{lcc}
\hline
                                               & $\Delta \reeta$ & $\Delta \imeta$ \\ \hline
Reconstruction efficiency        & $\pm \, 0.009$  & $\pm \, 0.013$  \\
$z_{\subrm{vertex}}$ resolution  & $\pm \, 0.010$  & $\pm \, 0.000$  \\
Beam geometry                    & $\pm \, 0.005$  & $\pm \, 0.007$  \\
Background                       & $\pm \, 0.002$  & $\pm \, 0.009$  \\
$\pid$ decays                    & $\pm \, 0.001$  & $\pm \, 0.001$  \\
Energy scale                     & $\pm \, 0.001$  & $\pm \, 0.001$  \\
Energy non-linearities           & $\pm \, 0.001$  & $\pm \, 0.002$  \\
$K^0 \overline{K^0}$ dilution    & $\pm \, 0.001$  & $\pm \, 0.001$  \\ \hline \hline
Total:                           & $\pm \, 0.015$  & $\pm \, 0.017$  \\
\hline
\end{tabular}
\caption{Systematic uncertainties on the $\etazzz$ measurement.}
\label{tab:systematics}
\end{center}
\spaceafterfloat
\end{table}

As a cross-check
we have fitted the ratio
$N_{K_L \to 3 \pi^0}^{\subrm{data}}(t)/N_{K_L \to 3 \pi^0}^{\subrm{MC}}(t)$
on the far-target data only.
We obtained $\reeta|_{\subrm{far-target}}=0.004 \pm 0.003$
and $\imeta|_{\subrm{far-target}}=-0.003 \pm 0.004$,
with the result being consistent with zero, as expected.

Two additional independent analyses have been performed,
yielding consistent results when applying similar cuts.
While one of these analyses employed a  method similar to the one described here~\cite{bib:ggthesis}, the other
analysis applied a toy Monte Carlo simulation for the geometry correction. 

\section{Results and Discussion}
\label{sec:result}

\vspace*{-5mm}

\begin{figure}[t]
\begin{center}
\epsfig{file=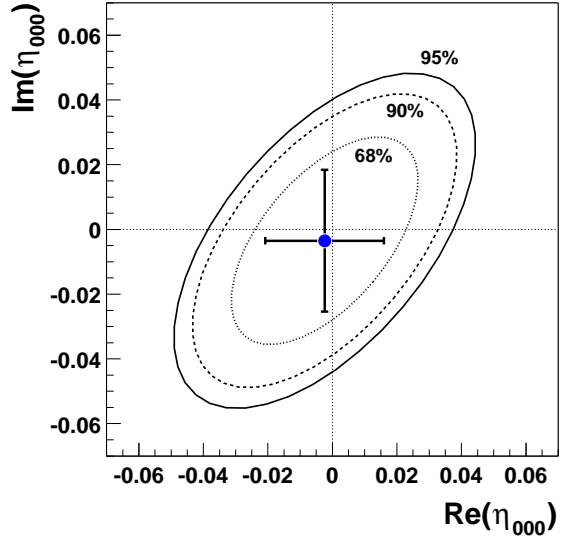,width=0.5\linewidth}
\caption{Fit result for $\etazzz$. The errors include both statistical and systematic uncertainties.
The lines indicate the exclusion limits for $68\%$, $90\%$, and $95\%$ confidence levels.}
\label{fig:resultellipse}
\end{center}
\spaceafterfloat
\end{figure}

Our result on $\etazzz$ is
\begin{equation}
\begin{array}{rcl}
\reeta & = & -0.002 \: \pm \: 0.011 \, {\rm (stat.)} \pm \: 0.015 \, {\rm (syst.)} \\
\imeta & = & -0.003 \: \pm \: 0.013 \, {\rm (stat.)} \pm \: 0.017 \, {\rm (syst.)} 
\end{array}
\label{eqn:result}
\end{equation}
with a statistical correlation coefficient of $0.77$
and an overall correlation coefficient of $0.57$ between real and imaginary parts (Fig.~\ref{fig:resultellipse}).
From this, an upper limit on the absolute value of $\etazzz$ is obtained as
\begin{equation}
|\etazzz| \; < \; 0.045
\end{equation}
at the 90\% confidence level.
Using the measured values of $\Br(\klpipipi) = 0.211 \pm 0.003$ and of the $\kl$ and $\ks$ lifetimes~\cite{bib:pdg},
this result turns into an upper limit on the $\kspipipi$ branching fraction of
\begin{equation}
\Br(\kspipipi) \; = \; |\etazzz|^2 \: \frac{\tau_S}{\tau_L} \: \Br(\klpipipi)
               \; < \; 7.4 \times 10^{-7}
\end{equation}
at a confidence level of $90\%$.
This value is more than one order of magnitude below the previous best limit~\cite{bib:ks3pi0_snd}.

Under the assumptions of CPT invariance
$\reeta \simeq {\rm Re}(\epsilon)$.
Fixing $\reeta = {\rm Re}(\epsilon) = 1.66 \times 10^{-3}$, the fit yields
$\imeta|_{CPT} = 0.000 \pm 0.009_{\subrm{stat}} \pm 0.013_{\subrm{syst}}$,
which can be turned into upper limits of
$|\etazzz|_{CPT} < 0.025$ and
${\rm Br}(\kspipipi)|_{CPT} < 2.3 \times 10^{-7}$
at a confidence level of $90\%$.

\begin{table}[t]
\begin{center}
\begin{tabular}{l@{\hspace*{1mm}}lrr}
$\alpha_f$          &                                     
                              & $10^3 \: \times$ Re$(\alpha_f)$  & $10^3 \: \times$ Im$(\alpha_f)$ \\ \hline
$\alpha_{+-}$       & $= \; \eta_{+-} \: \Br(\kspipi)$
                              & $1.146 \pm 0.015$                & $1.084 \pm 0.016$ \\ 
$\alpha_{00}$       & $= \; \eta_{00} \: \Br(\kspizpiz)$
                              & $0.511 \pm 0.008$                & $0.488 \pm 0.008$ \\ 
$\alpha_{+-\gamma}$ & $= \; \eta_{+-\gamma} \: \Br(\kspipig)$
                              & $0.003 \pm 0.000$                & $0.003 \pm 0.000$ \\ 
$\alpha_{l3}$       & $= \; 2 \: \frac{\tau_S}{\tau_L} \: \Br(\kl \to \pi l \nu) \: \times$ & & \\
                    & $\quad \; \; [{\rm Re}(\epsilon) - {\rm Re}(y) - i({\rm Im}(x_+) + {\rm Im}(\delta))]$
                              & $-0.001 \pm 0.007$              & $0.005 \pm 0.006$ \\ 
$\alpha_{+-0}$      & $= \; \frac{\tau_S}{\tau_L} \: \eta^\star_{+-0} \: \Br(\klpipipic)$
                              & $0.000 \pm 0.002$                & $0.000 \pm 0.002$ \\ 
$\alpha_{000}$      & $= \; \frac{\tau_S}{\tau_L} \: \eta^\star_{000} \: \Br(\klpipipi)$
                              & $-0.001 \pm 0.007$               & $0.001 \pm 0.008$ \\ \hline
$\sum \alpha_f$     &
                              & $1.658 \pm 0.024$               & $1.581 \pm 0.025$ \\ \hline
\end{tabular}
\caption{Parameters $\alpha_f = A(\kl \to f)^\star A(\ks \to f) / \Gamma_S$ for decays of neutral kaons
into the final state $f$. $\etazzz$ is taken from this measurement. The $\Delta S =\Delta Q$ violating parameter ${\rm Im}(x_+)$
and the CPT violating parameter ${\rm Re}(y)$ are taken from Ref.~\cite{bib:cpt_cplear}.
All other measurements are from Ref.~\cite{bib:pdg}.}
\label{tab:etas}
\end{center}
\spaceafterfloat
\end{table}

The result on $\etazzz$ from the fit without imposing CPT can be used to improve the 
test of CPT invariance via the Bell-Steinberger relation~\cite{bib:bellsteinberger}.
This relation uses unitarity to connect the CP violating amplitudes of $\ks$ and $\kl$ decays
with the CP violating parameter $\epsilon$ and 
the CPT violating parameter $\delta$ through
\begin{equation}
\begin{array}{rcl}
(1 + i \tan{\phi_{SW}}) [ {\rm Re}(\epsilon) - i \: {\rm Im}(\delta) ]
               & = & {\displaystyle{\sum_{{\subrm{final}} \atop {\subrm{states}} \: f}}} A(\kl \to f)^\star A(\ks \to f) / \Gamma_S \\*[1mm]
               & = & {\displaystyle{\sum_{{\subrm{final}} \atop {\subrm{states}} \: f}}} \alpha_f
\end{array}
\label{eqn:cpt}
\end{equation}
where the super-weak angle is defined by 
$\tan{\phi_{SW}} = 2 \, (m_L - m_S)/(\Gamma_L - \Gamma_S)$.
Up to now the limit on Im$(\delta)$ was limited by poor knowledge of $\etazzz$.
Together with available measurements of the other CP violating amplitude ratios (see Table~\ref{tab:etas}), the $\kl$ and $\ks$ lifetimes, 
branching fractions, and
mass difference~\cite{bib:pdg}, and the parameters ${\rm Re}(y)$ and ${\rm Im}(x_+)$ 
of possible CPT and $\Delta S = \Delta Q$ violation~\cite{bib:cpt_cplear}, 
the new result presented here (Eq.~\ref{eqn:result}) can be used
to improve the knowledge of ${\rm Im}(\delta)$ and ${\rm Re}(\epsilon)$.
Solving the complex equation (\ref{eqn:cpt}) for the free parameters ${\rm Im}(\delta)$ and ${\rm Re}(\epsilon)$,
taking into account correlations between input parameters, yields
\begin{equation}
\begin{array}{rcl}
{\rm Im}(\delta)   & = & (  -0.2 \, \pm \, 2.0 ) \times 10^{-5}, \\
{\rm Re}(\epsilon) & = & ( 166.4 \, \pm \, 1.0 ) \times 10^{-5}.
\end{array}
\end{equation}
The errors on both ${\rm Im}(\delta)$ and ${\rm Re}(\epsilon)$ are reduced by a factor of $2.5$ with respect to the
previous values~\cite{bib:cpt_cplear} and are now limited by knowledge of $\eta_{+-}$.

Finally, when assuming CPT invariance in the decay, this can be converted
into a measurement of the $K^0 \overline{K^0}$ mass difference
of
\begin{equation}
m_{K^0} - m_{\overline{K^0}} \: = \: (-0.2 \, \pm \, 2.8) \times 10^{-19} \: {\rm GeV}/c^2,
\end{equation}
and yields an upper limit of
$| m_{K^0} - m_{\overline{K^0}} | \: < \: 4.7 \times 10^{-19} \: {\rm GeV}/c^2$
at the 90\% confidence level,
which is the most precise test of CPT invariance in the mass matrix.

\section{Acknowledgements}

\vspace*{-5mm}

It is a pleasure to thank the technical staff of the participating
laboratories, universities, and affiliated computing centres for their
efforts in the construction of the NA48 apparatus, in the
operation of the experiment, and in the processing of the data.

\end{document}